\documentclass[aps,prd,nofootinbib,twocolumn,superscriptaddress,preprintnumbers]{revtex4}

\usepackage{amssymb}
\usepackage{amsmath}
\usepackage{epsfig}
\usepackage{hyperref}
\usepackage{breakurl}

\usepackage{undertilde}
\makeatletter
\def\simgt{\mathrel{\lower2.5pt\vbox{\lineskip=0pt\baselineskip=0pt
           \hbox{$>$}\hbox{$\sim$}}}}
\def\simlt{\mathrel{\lower2.5pt\vbox{\lineskip=0pt\baselineskip=0pt
           \hbox{$<$}\hbox{$\sim$}}}}
\makeatother

\newcommand{\be}{\begin{equation}}
\newcommand{\ee}{\end{equation}}
\newcommand{\bea}{\begin{eqnarray}}
\newcommand{\eea}{\end{eqnarray}}
\newcommand{\Eq}[1]{Eq.~(\ref{#1})}

\newcommand{\Fig}[1]{Fig.~(\ref{#1})}

\newcommand{\LL}{\mathcal{L}}
\newcommand{\OO}{\mathcal{O}}
\newcommand{\BR}{\textrm{BR}}

\begin{document}

\title{Seesaw Spectroscopy at Colliders}

\author{Clifford Cheung}
\affiliation{Berkeley Center for Theoretical Physics, 
  University of California, Berkeley, CA 94720, USA}
\affiliation{Theoretical Physics Group, 
  Lawrence Berkeley National Laboratory, Berkeley, CA 94720, USA}

\author{Lawrence J. Hall}
\affiliation{Berkeley Center for Theoretical Physics, 
  University of California, Berkeley, CA 94720, USA}
\affiliation{Theoretical Physics Group, 
  Lawrence Berkeley National Laboratory, Berkeley, CA 94720, USA}
\affiliation{Institute for the Physics and Mathematics of the Universe, 
  University of Tokyo, Kashiwa 277-8568, Japan}
  
  \author{David Pinner}
\affiliation{Berkeley Center for Theoretical Physics, 
  University of California, Berkeley, CA 94720, USA}
\affiliation{Theoretical Physics Group, 
  Lawrence Berkeley National Laboratory, Berkeley, CA 94720, USA}

\begin{abstract}

A low-scale neutrino seesaw may be probed or even reconstructed at colliders provided that supersymmetry is at the weak scale and the LSP is a sterile sneutrino.  Because the neutrino Yukawa couplings are small, the NLSP is typically long-lived and thus a significant fraction of colored or charged NLSPs may stop in the detector material before decaying to the LSP and a charged lepton, gauge boson, or Higgs.   For two-body NLSP decays, the energy spectrum of the visible decay product exhibits a monochromatic line for each sterile sneutrino which can be used to extract the sterile sneutrino masses and some or all entries of the neutrino Yukawa matrix modulo phases.  Similar methods can be used to extract these parameters from the Dalitz plot in the case of three-body NLSP decays.  Assuming that the sterile sneutrino and neutrino are roughly degenerate, one can confirm the existence of a neutrino seesaw by comparing these measured parameters to the observed active neutrino masses and mixing angles.  Seesaw spectroscopy can also provide genuinely new information such as the value of $\theta_{13}$, the nature of the neutrino mass hierarchy, and the presence of CP conservation in the neutrino sector.  We introduce a weak-scale theory of leptogenesis that can be directly tested by these techniques.
 

\end{abstract}

\maketitle

\section{Introduction}

The seesaw mechanism \cite{Yanagida:1979as} offers an exceptionally elegant explanation for the origin of neutrino masses.  In its minimal incarnation---the so-called type I seesaw---the standard model is supplemented by three generations of sterile neutrinos, $\nu^c_i$, which couple via
\bea
\LL &=&  \lambda_{ij}\ell_i   \nu^c_j h +  \frac{1}{2} M_{i} \nu^c_i  \nu^c_i.
\eea
At energies well below the sterile neutrino masses, $M_i$, and the electroweak symmetry breaking scale, $v$, the active neutrinos acquire masses, $m_i$, which are the eigenvalues of the matrix
\bea
m_{ij} &=& v^2 (\lambda M^{-1} \lambda^T)_{ij}.
\eea
The active neutrino masses are stringently bounded by cosmological measurements \cite{Seljak:2006bg} 
\bea
\sum_{i} m_i < 0.17\, \mbox{eV}, 
\label{eq:neutrinomassbound}
\eea
while experimental measurements of solar, atmospheric, accelerator and reactor neutrinos constrain the neutrino mass differences and mixing angles \cite{Nakamura:2010zzi},
\bea
|\Delta m_{32}^2| &=& (2.43 \pm 0.13)\cdot 10^{-3} \textrm{ eV}^2 \nonumber \\
|\Delta m_{21}^2| &=& (7.59 \pm 0.20)\cdot 10^{-5} \textrm{ eV}^2 \nonumber \\
\sin^2 2\theta_{12} &=& 0.87 \pm 0.03 \nonumber \\
\sin^2 2 \theta_{23} &>& 0.92 \nonumber \\
\sin^2 2\theta_{13} &<& 0.15, 
\label{eq:neutrinoprops}
\eea
where $\Delta m^2_{ij} \equiv m_i^2 - m_j^2$.
 In the high-scale seesaw, $M_i\simeq 10^{15}$ GeV, allowing for order unity Yukawa couplings, 
 $\lambda_{ij} \simeq \OO(1)$.  On the other hand, the sterile neutrinos can also be much lighter, $M_i\lesssim 100$ GeV, if one is willing to accommodate markedly smaller Yukawa couplings, $\lambda_{ij} \lesssim \OO(10^{-6})$.  Considering that the electron Yukawa is $\OO(10^{-5})$, the low-scale seesaw is not terribly unreasonable.  Small Yukawas notwithstanding, there is still a substantial ``seesaw effect'' at work in this regime, since the active neutrino masses are hierarchically smaller than the sterile neutrino masses.  Furthermore, although thermal leptogenesis typically requires $M_i \gtrsim 10^9$ GeV, it is nevertheless possible to generate the observed baryon asymmetry in other models with $M_i$ near the weak scale \cite{Murayama:1993em, Pilaftsis:2003gt, Grossman:2004dz}.

A direct probe of the seesaw mechanism at colliders is considered impossible because the sterile neutrinos are either incredibly heavy or incredibly weakly coupled to the standard model.  As a consequence, the seesaw is typically constrained by very indirect probes, for example testing the Majorana nature via measurements of neutrinoless double $\beta$ decay or lepton flavor violation, such as $\mu \rightarrow e \gamma$, in models with weak-scale supersymmetry.

In this paper, we show that the seesaw may be directly probed and in some cases {\it reconstructed} at colliders if there is weak-scale supersymmetry and the LSP is a sterile sneutrino.  The superpotential is given by
\bea
W &=& \lambda_{ij} L_i N^c_j H_u + \frac{1}{2} M_{i} N^c_i N^c_i ,
\eea
where $M_i$ is at or below the weak-scale in order to accommodate a sterile sneutrino LSP.   Throughout, we assume that supersymmetry breaking in the $N^c_i$ sector is highly suppressed, as occurs in gauge mediation\footnote{In this case the sneutrino decays to the true LSP, the gravitino, but as we will see our analysis will be unaffected by this point.}, so the sterile neutrinos and sneutrinos are essentially degenerate.

Consider, for the sake of example, the case of chargino NLSP\footnote{While chargino NLSP is difficult to achieve in the MSSM \cite{Kribs:2008hq}, we consider it here for concreteness.}, which decays to a lepton and a sterile sneutrino LSP via $\tilde\chi^\pm \rightarrow \ell^\pm_i \tilde \nu^c_j$.
The chargino lifetime, along with the nine branching ratios 
\bea \BR(\tilde\chi^\pm \rightarrow \ell^\pm_i \tilde \nu^c_j) \propto |\lambda_{ij}|^2
\eea 
are in direct correspondence with the magnitudes of the nine entries of the Yukawa coupling matrix.  Furthermore,
the sterile sneutrino masses are directly related to the chargino mass and the energy of the outgoing lepton in the chargino rest frame through
\bea
M_{j}^2 = m_{\tilde \chi^\pm}^2 + m_{\ell^\pm_i}^2 - 2 m_{\tilde \chi^\pm} E_{\ell^\pm_i}.
\label{eq:massrelation}
\eea
Because $|\lambda_{ij}|$ and $M_j$ are related to physical quantities that can, in principle, be measured directly, we are left with the remarkable prospect of actually reconstructing the origin of neutrino masses at colliders.  At the order of magnitude level, the seesaw may be verified by measuring
\bea
\frac{\lambda^2 v_u^2}{M} \sim 0.1\, \mbox{eV},
\eea
in which $\lambda$ and $M$ are representative values of the Yukawa coupling matrix and the sterile sneutrino masses.  As we will see, similar statements can be made for squark, slepton, and gluino NLSPs.


In what follows, we evaluate to what extent this idealized proposal might be realistically achieved at the LHC.  Given existing measurements of the active neutrino masses, the decay of the NLSP to a sterile sneutrino LSP is typically displaced ($\gamma c \tau \gtrsim 1 \textrm{ mm}$) or even long-lived ($\gamma c \tau \gtrsim 1 \textrm{ m}$) at colliders.  In the latter case, metastable charged or colored NLSPs  may actually be produced in supersymmetric cascades and stopped in the ATLAS or CMS detector apparatus \cite{Khachatryan:2010uf, Asai:2009ka, Arvanitaki:2005nq}.  Precision measurements can then be employed to determine $|\lambda_{ij}|$ from the lifetime and branching ratios of the NLSP, while $M_i$ may be ascertained from energy spectrum of the decay products.  We dub this procedure seesaw spectroscopy.
Similar techniques have in the past been proposed as a sensitive probe of split supersymmetry \cite{Arvanitaki:2005nq} as well as very weakly coupled states like gravitinos \cite{Buchmuller:2004rq}, axinos \cite{Brandenburg:2005he}, goldstini \cite{Cheung:2010mc,Cheung:2010qf}, and dark matter \cite{Feng:2004mt,Cheung:2010gk}.  Our analysis will address how stopped NLSPs might allow for the seesaw mechanism to be  verified at collider experiments.

In Section 2 we discuss the conditions under which the low-scale seesaw might be verified or excluded given measurements of the neutrino Yukawa couplings and the sterile sneutrino masses.  We go on to discuss how these parameters might be realistically obtained from collider experiments in Section 3.  Finally, in Section 4 we conclude with a discussion of a remarkable version of weak scale leptogenesis---asymmetric freeze-in---in which the neutrino sector parameters relevant for the lepton asymmetry are directly accessible via seesaw spectroscopy. 

\begin{figure}[t]
\begin{center} 
\includegraphics[scale=0.4]{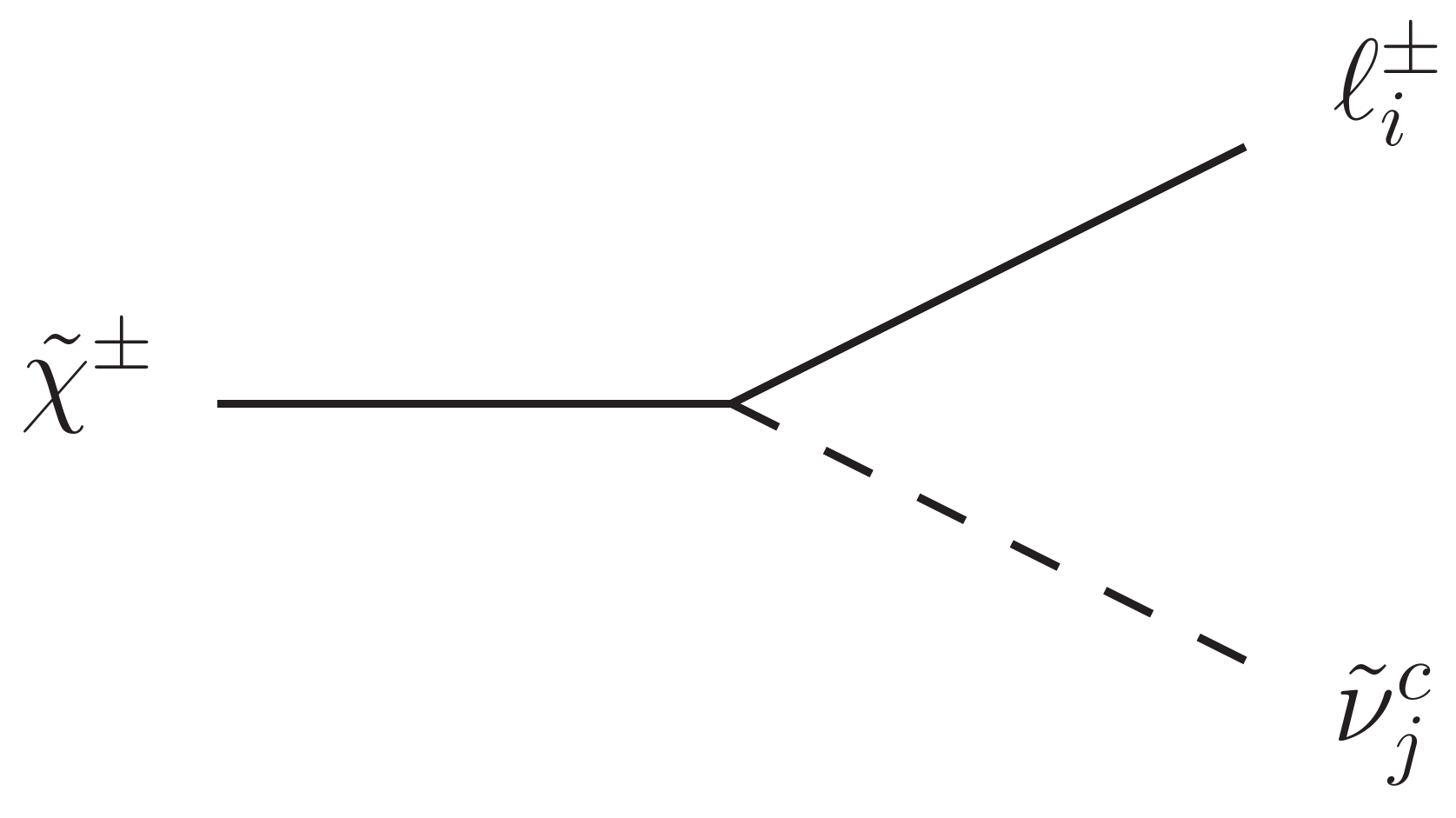}
\end{center}
\caption{The lifetime and branching ratios associated with the chargino decay $\tilde \chi^\pm \rightarrow \ell^\pm_i \tilde \nu^c_j$ can be used to determine $|\lambda_{ij}|$, while the lepton spectrum can be used to reconstruct $M_j$. }
\label{fig:decay}
\end{figure}

\section{Seesaw Signals}

Due to  R-parity the NLSP is required to have some non-zero decay width to sterile sneutrinos via the seesaw Yukawa couplings.   In this section we show how existing neutrino experiments can provide precise correlations which relate the NLSP lifetime and branching ratios to the sneutrino masses.  For the moment, let us assume that these quantities can be successfully measured at colliders.  We will show that this assumption is reasonable in Section 3.

The decay width of a pure charged higgsino or left-handed slepton NLSP into sterile sneutrinos is given, up to phase space factors, by
\bea
\label{eq:gammas}
\Gamma(\tilde \chi^\pm \rightarrow  \ell^\pm\tilde \nu^c)  &\simeq&\frac{ m_{\tilde \chi^\pm}}{32\pi}  \sum_{ij} |\lambda_{ij}|^2 \\
\Gamma(\tilde \ell^\pm_i \rightarrow  W^\pm \tilde\nu^c)  &\simeq& \frac{1}{{16\pi m_{\tilde \ell^\pm}}}\sum_{j}  |\lambda_{ij} (M_j \sin \beta - \mu^* \cos \beta)|^2  \nonumber \\
\Gamma(\tilde \ell^\pm_i \rightarrow  h^\pm \tilde\nu^c)  &\simeq& \frac{1}{{16\pi m_{\tilde \ell^\pm}}}\sum_{j}  |\lambda_{ij} (M_j \cos \beta+ \mu^* \sin \beta)|^2,  \nonumber
\eea
where the slepton decay occurs through the supersymmetric trilinear Yukawa couplings, and $j$ implicitly sums over kinematically accessible sneutrinos.  In any realistic scenario, the NLSP will of course include some admixture of states, in which case the decay width of the corresponding chargino or slepton will be dressed by the relevant mixing angles.  Likewise, if the NLSP is a squark or gluino, then its decay will be mediated by an off-shell charged higgsino, and so its decay width will be suppressed by additional phase space factors and ratios of particle masses.  

\subsection{Discovering the Seesaw}

The above formulae immediately allow for the following simple-minded, order of magnitude check of the neutrino seesaw at colliders.  The lifetime, $\tau = \Gamma^{-1}$,  and branching ratios of the NLSP yield an indirect measurement of the entries of $|\lambda_{ij}|$.  Moreover, $M_i$ can in principle be measured from the energy spectrum of the NLSP decay products.  For the case of pure charged higgsino NLSP,
\bea
c \tau \sim 1 \textrm{ mm} \times \sin^2 \beta  \left(\frac{150 \text{ GeV}}{m_\chi}\right)\left(\frac{0.1 \text{ eV}}{m}\right)\left(\frac{10 \text{ GeV}}{M}\right) 
\label{eq:longtau}
\eea
so the decay length can be quite long.  Note that for this order of magnitude confirmation we have inserted degenerate values for $\lambda_{ij} = \lambda \delta_{ij}$,  $M_i = M$, and $m_i = m$ such that $\lambda^2 v_u^2 / M = m \sim 0.1 \textrm{ eV}$, which is the natural scale expected for the active neutrino masses given \Eq{eq:neutrinomassbound} and \Eq{eq:neutrinoprops}.  This number is roughly the same for a pure left-handed slepton NLSP, but can be substantially greater if one includes additional mixing angles or if the NLSP is a squark or gluino.  For example, gauge mediation typically results in right-handed slepton NLSPs, enhancing the slepton NLSP lifetime by a factor of $m_{\tilde{\ell}^{\pm}}^2/m_{\ell^{\pm}}^2$ once left-right mixing is included.  These long lifetimes allow for stopping charged or colored NLSPs in the detector, improving the reconstructability of the sterile sneutrino masses from kinematic measurements.

Observing this order of magnitude relationship would provide substantial evidence for the seesaw as the origin of neutrino masses.  That said, there are a number of important subtleties which one must address even in this simplest litmus test.  

For example, the NLSP may have additional decay channels beyond the sneutrino which dominate the full width.  This is true in very low-scale gauge mediated supersymmetry breaking, $\sqrt{F} \lesssim 100$ TeV, where the NLSP has a dominant decay width to gravitinos.  In this scenario the overall scale of $\lambda_{ij}$ is naively inaccessible because $\tau$ is fixed by the supersymmetry breaking scale, $\sqrt{F}$, which is unrelated to the seesaw.  Nonetheless, one can still measure the branching ratio of sneutrinos versus gravitinos, in which case the overall scale of $\lambda_{ij}$ can still be inferred and the above analysis applies.  Alternatively, decays to gravitinos can easily be sub-dominant, requiring only $\sqrt{F} \gtrsim 100$ TeV.

The story is also more complicated if the NLSP is a slepton.  Typically, in this case only a single row of $\lambda_{ij}$ is accessible from slepton NLSP decays, since the generational index of the NLSP is fixed.  However, some theories possess sufficiently mass degenerate slepton co-NLSPs \cite{Giudice:1998bp}, in which case more information may be available.

In addition to the correlation between the active neutrino mass scale and the NLSP lifetime, further evidence for the seesaw may be discovered by measuring the branching ratios of NLSP decays to each of the lepton flavors.  The existence of large neutrino mixing angles suggests that one should measure roughly equal parts $e$, $\mu$, and $\tau$ arising from each sneutrino in the NLSP decays.



\subsection{Probing the Seesaw}

Checking the order of magnitude validity of \Eq{eq:longtau} would be an important first step towards performing precision measurements of the seesaw mechanism at colliders.  Of course, this coarse analysis disregards all of the flavor information which  can be extracted from a measurement of the full $|\lambda_{ij}|$ matrix and the sneutrino masses $M_{i}$.  Indeed, all seesaw parameters are accessible to experiment except for 1) the phases of the Yukawa couplings, $\arg \lambda_{ij}$, and 2) any Yukawa couplings or masses related to kinematically inaccessible sneutrinos.  In the following section, we discuss how additional flavor information can provide essential data about the neutrino seesaw.

The measured values of $|\lambda_{ij}|$ and $M_i$ can be combined to form an ``effective'' neutrino mass matrix
\bea 
\hat m_{ij} &=& v_u^2 \sum_k  |\lambda_{ik}| \frac{1}{M_k} |\lambda^T_{kj}| ,
\label{eq:mhat}
\eea
where $k$ sums over all sneutrinos lighter than the NLSP and $\tan\beta$ is assumed to be measured.  Clearly, $\hat m_{ij}$ is missing all information in the theory involving phases or kinematically inaccessible sneutrinos.  In general, $\hat m_{ij} \neq m_{ij}$, the actual neutrino mass matrix, and consequently the neutrino masses and mixing angles derived from $\hat m_{ij}$ will not match the observations of existing neutrino experiments shown in \Eq{eq:neutrinomassbound} and \Eq{eq:neutrinoprops}.  However, in many cases the eigenvalues and mixing angles associated with $\hat m_{ij}$ may be found to match $m_{ij}$.  If this occurs, then the information provided in $\hat m_{ij}$ can provide new and substantive information about the neutrino sector.  

For example, if all three sneutrinos are kinematically accessible from the NLSP decay and one discovers that $\hat m_{ij} = m_{ij}$ identically, then this would indicate that the CP phases in the neutrino sector are negligible.  This would be strong evidence that the neutrino Yukawa couplings respect CP symmetry.  If this were true, then colliders could provide unambiguous information about as of yet unknown quantities like $\theta_{13}$ or the nature of the active neutrino mass hierarchy!

Alternatively, suppose that the heaviest active neutrino mass is generated by integrating out the lightest sterile neutrino, $\nu^c_1$.  If the lightest sneutrino is the only sterile neutrino accessible in NLSP decay, so that only $k=1$ contributes in \Eq{eq:mhat}, then the NLSP lifetime will be precisely correlated with $m_{\mbox{atm}} = \sqrt{|\Delta m_{32}^2|}$.  Furthermore, the ratio of $\tau:\mu:e$ events will be $0.5:0.5:\theta_{13}^2$.  In this case, the phases in the neutrino Yukawa couplings may be rotated into the elements corresponding to the heavier sneutrinos, so that $\hat m_{ij} \approx m_{ij}$, up to corrections of $\OO(m_{\mbox{sol}}/m_{\mbox{atm}})$ where $m_{\mbox{sol}} = \sqrt{|\Delta m_{21}^2|}$.  Such measurements would convincingly demonstrate the weak-scale seesaw origin of neutrino masses and provide a measurement of the presently unknown angle $\theta_{13}$.


\subsection{Excluding the Seesaw}

Collider measurements can be used to exclude the seesaw as well.  For instance, consider a scenario in which the NLSP can decay to all three sterile sneutrinos.  For concreteness, we assume a pure charged higgsino NLSP, although our discussion is easily generalized to arbitrary NLSP.   The decay width may be rewritten as
\bea
\Gamma(\tilde\chi^\pm \rightarrow \ell^\pm \tilde \nu^c) 
 &=& \frac{ m_{\tilde \chi^\pm}}{32\pi} \sum_{i}\lambda_{i}^2,
\label{eq:partwidth}
\eea
where $\lambda_i$ are the real and positive semi-definite diagonal entries of the diagonalized Yukawa interaction matrix, $(L \lambda R^\dagger)_{ij}$.
The active neutrino masses are given by
\bea
\sum_i m_i^2 &=& v_u^4 \sum_{ijkl} \lambda_i^2 R^\dagger_{ij} \frac{1}{M_j} R^*_{jk} \lambda^2_k R^T_{kl} \frac{1}{M_l} R_{li} \\
 &\leq& v_u^4 \sum_{ijkl} \lambda^2_i \frac{1}{M_j} \lambda^2_k \frac{1}{M_l} \\
 &=& v_u^4 (\sum_i \lambda_i^2)^2 (\sum_j M^{-1}_j)^2,
\eea
where the inequality in the second line arises from the fact that every entry of a unitary matrix is bounded by unity, so $|R_{ij}|<1$.  Using this expression in \Eq{eq:partwidth} yields
\bea
\frac{32 \pi \Gamma}{m_{\tilde \chi^\pm}} &\geq& \frac{ \sqrt{\sum_i m_i^2}  (\sum_j M_j^{-1})^{-1}}{v_u^2}.
\eea
Since the sum of active neutrino masses is bounded from below by measurements of $|\Delta m_{32}^2|$ and the partial decay width of the chargino NLSP to sterile sneutrinos in \Eq{eq:partwidth} is constrained from above by the inverse lifetime, $\tau^{-1}$, we obtain our final bound,
\bea
(\sum_i M_i^{-1})^{-1} &\leq& 81 \textrm{ GeV} \times\sin^2 \beta \times
\left(\frac{200 \textrm{ GeV}}{m_{\tilde \chi^\pm}}\right) \left(\frac{1 \textrm{ mm}}{c\tau}\right). \nonumber
\label{eq:taubound}
\eea
If the sneutrino masses are successfully measured and this inequality fails, then the physics being probed is not the origin of neutrino masses.  Analogous bounds to \Eq{eq:taubound} can of course be derived for alternative NLSP candidates.





\begin{figure}[t]
\begin{center} 
\includegraphics[scale=1]{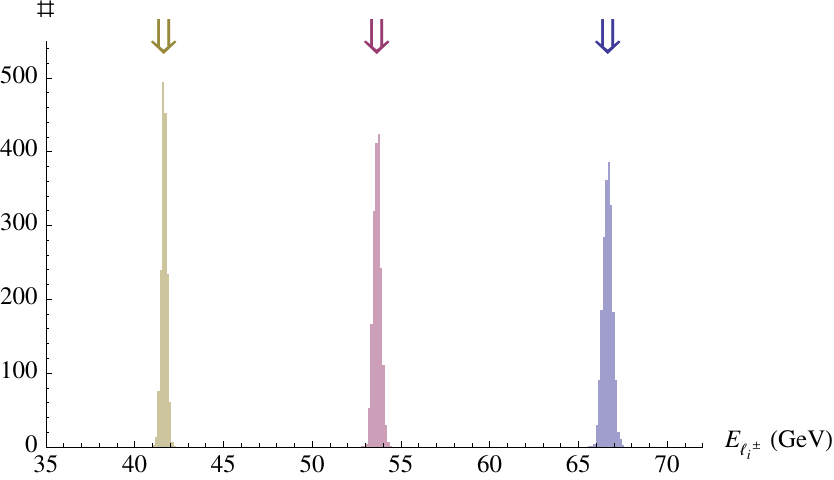}
\end{center}
\caption{Histogram of the lepton energy from $\tilde \chi^\pm \rightarrow \ell^\pm_i \tilde \nu_j^c$ for $m_{\tilde \chi^\pm} =$ 150 GeV.  The \{purple, red, yellow\}  peaks correspond to the sterile sneutrinos at masses $M_j =$ \{50, 80, 100\} GeV and the arrows correspond to the values predicted by \Eq{eq:massrelation}.  We have taken $\delta E_{\ell^\pm_i} = 0.004 E_{\ell^\pm_i}$.
}
\label{fig:2body}
\end{figure}

\section{Seesaw Spectroscopy }

In this section we evaluate to what extent the NLSP lifetimes, branching ratios, and sneutrino masses might be realistically  measured at the LHC.
Assuming that the dominant decay channel of the NLSP is to sterile sneutrinos, then \Eq{eq:longtau} implies that the NLSP can easily be long-lived when one includes various mixing angles and kinematic factors.   If the NLSP is charged or colored, then some fraction of them will be stopped inside the ATLAS or CMS detectors.
 
It has been shown \cite{Arvanitaki:2005nq} that upwards of $10^6$ long-lived, colored particles may be stopped each year at the LHC.  Indeed, searches for such particles have already been performed \cite{Khachatryan:2010uf}, using dedicated triggers which detect decays occurring out of time with the beam.  The aforementioned analysis considered the dominant interactions between the NLSP and the detector to be electromagnetic in nature, with the NLSP energy loss described by the Bethe-Bloch equation, and so it is equally applicable to both slepton NLSPs and squarks or gluinos which hadronize into charged particles.  Since these squarks or gluinos must decay through an intermediate chargino, they will decay to a three-body final state, extending their lifetime by a three-body phase space factor and an off-shell propagator.  As discussed above, slepton and chargino NLSPs will typically have their lifetimes extended by left-right mixing and higgsino-wino mixing, respectively.  Thus, we expect NLSP lifetimes to be sufficiently long in many cases to produce substantial numbers of stopped particles.
 
The exact number of stopped NLSPs is strongly dependent on the lifetime of the NLSP, which is in turn fixed by the nature of the NLSP and the mixing angles in the supersymmetric spectrum.  In order to sidestep this model dependence, we will divide the remainder of our discussion according to whether a sizeable fraction of NLSPs  is stopped or not.

\begin{figure}[t]
\begin{center} 
\includegraphics[scale=.8]{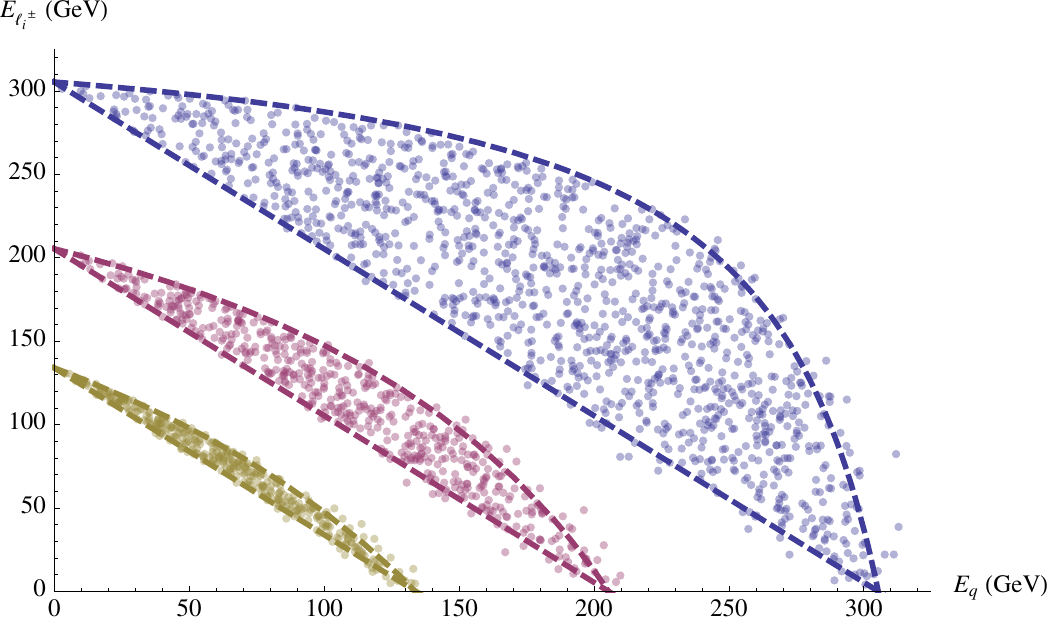}
\end{center}
\caption{Scatter plot of the quark and lepton energies from $\tilde q \rightarrow q \ell^\pm_i \tilde \nu_j^c$ for $m_{\tilde q} =$ 700 GeV.  The \{purple, red, yellow\}  points correspond to the sterile sneutrinos at masses $M_j =$ \{250, 450, 550\} GeV and the dashed lines correspond to endpoints analytically determined by the kinematics.  We have taken $\delta E_q = 0.033 E_q$ and $\delta E_{\ell^\pm_i} = 0.004 E_{\ell^\pm_i}$.
}
\label{fig:3body}
\end{figure}

\subsection{Stopped NLSPs}

A large number of NLSPs will be stopped at LHC if the NLSP lifetime is sufficiently long.  In this case the ensuing decays occur in the rest frame of the detector, and so the initial four-vector of the NLSP is known, assuming the NLSP mass has been measured by other means.  Additionally, since decays from stopped particles occur out of time with respect to the main collision events, we expect signals from such decays to be exceptionally clean.

If the NLSP decay is two-body then the visible decay product will be monochromatic with an energy uniquely fixed by the NLSP mass and the sneutrino mass, as shown for the case of chargino NLSP in \Eq{eq:massrelation}.  In this case each sneutrino decay product will produce a separate peak in the energy spectrum of the outgoing lepton.

In many cases it should be easy to distinguish the peaks corresponding to each sneutrino, even assuming uncertainties in the energy measurement of the outgoing charged lepton, gauge boson, or Higgs.  \Fig{fig:2body} shows a histogram of the energy spectrum of the outgoing lepton\footnote{Representative values for the energy uncertainties in \Fig{fig:2body} and \Fig{fig:3body} are taken from the ATLAS TDR.} arising from the decay of a chargino NLSP, $\tilde \chi^\pm \rightarrow \ell^\pm_i \tilde \nu^c_j$.  By measuring the central value of each peak, one can reconstruct the mass of each corresponding sneutrino.  Likewise, by counting the total number of events within each peak and tagging the flavor of $\ell_i^\pm$, one obtains the branching ratio $\textrm{BR}(\tilde \chi^\pm \rightarrow \ell^\pm_i \tilde \nu^c_j)$, and hence the value of each entry of $|\lambda_{ij}|$.  

Due to the small uncertainty in the lepton energy resolution, the dominant error in measuring the sneutrino mass by this method comes from uncertainty in the NLSP mass, $\delta M_j  \approx \delta m_{\tilde{\chi}^{\pm}}(m_{\tilde{\chi}^{\pm}} - E_{\ell^{\pm}_i})/M_j$.  Additionally, cosmic rays provide a significant background to these events.  As a result, current searches veto on reconstructed muons in the final state; however, for the purposes of reconstructing the seesaw, it will be necessary to distinguish cosmic ray muons from muonic NLSP decays.

If the NLSP decays to a three-body final state, then the analysis is slightly more involved.  For example, consider the case of a three-body decay of a squark NLSP, $\tilde q \rightarrow q \ell_i^\pm \tilde \nu^c_j$, via an off-shell chargino.  While the quark and lepton energies are not monochromatic, they are constrained by kinematic endpoints dictated by the mass of the squark and sneutrino.  For example, \Fig{fig:3body} depicts a scatter plot of quark and lepton energies from the squark NLSP decay.  As expected, one obtains three distinct bands corresponding to the decay of the squark to each sneutrino.  As in the two-body case, it should be possible to distinguish between these bands for a broad range of theories.  If this is the case, then the shape and location of each band can be used to determine the mass of the corresponding sneutrino, and the number of events in each band along with the flavor of the outgoing lepton can provide the associated branching ratio, $\textrm{BR}(\tilde q \rightarrow q \ell^\pm_i \tilde \nu^c_j)$.  Note that even in the case where the bands overlap, it may still be possible to distinguish them since there is additional information in the number of events in each band.  For example, for a constant matrix element, the bands will appear as three distinct plateaus, each sitting atop the next.

\begin{figure}[t]
\begin{center} 
\includegraphics[scale=0.75]{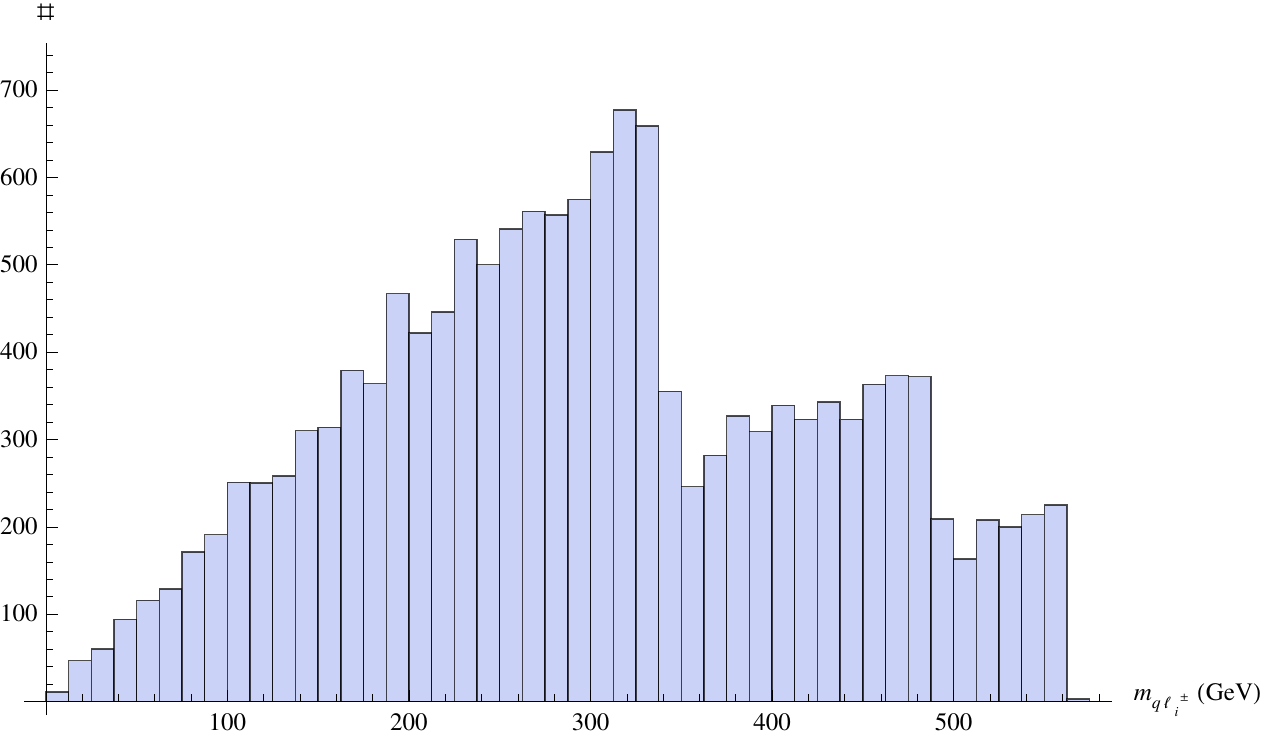}
\end{center}
\caption{Histogram of the invariant mass of the outgoing quark and lepton from $\tilde q \rightarrow  q \tilde\chi^\pm \rightarrow q \ell^\pm_i \tilde \nu_j^c$ for $m_{\tilde q} =$ 600 GeV and $m_{\tilde \chi^\pm}=$ 200 GeV.  The edges correspond to sterile sneutrinos at masses $M_j =$ \{20, 100, 160\} GeV.}
\label{fig:edge}
\end{figure}

\subsection{Non-Stopped NLSPs}

If the NLSP is too short-lived to be efficiently stopped, then it will typically be moving when it decays.  In this scenario the initial four-vector of the NLSP is unknown, so event reconstruction is a greater challenge.   Nonetheless, if the NLSP arises from a cascade decay, then kinematical information from the entire cascade can be used to reconstruct the sneutrino masses.  

For example, for a squark to chargino to sneutrino cascade decay,  $\tilde q \rightarrow  q \tilde\chi^\pm \rightarrow q \ell^\pm_i \tilde \nu_j^c$, there is a distinct edge in the invariant mass constructed from the quark and lepton which occurs at
\bea
m_{q\ell^\pm_i}^2 &=& \frac{(m_{\tilde q}^2-m_{\tilde \chi^\pm}^2)(m_{\tilde \chi^\pm}^2-M_j^2)}{m_{\tilde \chi^\pm}^2}.
\eea
As shown in \Fig{fig:edge}, each sneutrino edge may be visible in the $m_{q\ell^\pm_i}$ spectrum and it may be possible to reconstruct the branching ratios of the NLSP as well as the masses of each sneutrino.  

Nevertheless, there is some subtlety required in measuring the $m_{q\ell^\pm_i}$ spectrum.  Since each chargino decay is displaced, one can efficiently identify the associated leptons by selecting for hard leptons displaced from the primary vertex.  On the other hand, there are intrinsic combinatoric jet backgrounds which remain even after cutting on soft jets in the event.  If the NLSP decays are truly two-body, it may be possible to mitigate problems associated with jet combinatorics by considering only the lepton spectrum and the missing transverse momentum.  Techniques based on the $M_{T2}$ kinematic variable \cite{Konar:2009wn} could provide a measurement of the sneutrino mass independent of both the jet kinematics and the NLSP mass in cases where the NLSP is pair produced. Note that because these measurements do not require stopped NLSPs, they apply to the left-handed sneutrino NLSP as well as charged and colored NLSPs.

\section{Conclusion}

In this paper we have discussed the possibility of seesaw spectroscopy at colliders: a procedure whereby detailed information concerning the masses and mixing angles of the neutrino sector might be directly extracted from the decays of long-lived NLSPs.  Given the robustness of this probe, this proposal naturally begs the question: might seesaw spectroscopy be used as a tool to study the origin of the cosmological baryon asymmetry via leptogenesis?  

Such an ambitious endeavor is not possible in many theories, including standard high-scale thermal leptogenesis, since $M_i \gtrsim 10^9$ GeV \cite{Davidson:2002qv}. Soft leptogenesis via CP violating mixing of sterile sneutrinos allows significantly lower values of $M_i$, but still very much larger than the weak scale \cite{Grossman:2003jv,D'Ambrosio:2003wy}.  Soft leptogenesis with $M_i$ of order the weak scale is possible via CP violating decays of sterile sneutrinos \cite{Grossman:2004dz}, and weak scale leptogenesis can also result from the decay of sterile sneutrino condensate formed during inflation \cite{Murayama:1993em}.  However, in both these cases the LSP is a standard model superpartner and not a sterile sneutrino, so there can be no seesaw spectroscopy signals of long-lived NLSP decays to sterile sneutrinos.  Resonant leptogenesis can occur at the weak scale \cite{Pilaftsis:2003gt} but the size of the asymmetry is linked to the very high degeneracy of the $M_i$ which again cannot be probed via seesaw spectroscopy, even if the theory is supersymmetric. 

On the other hand, a reconstructable theory of leptogenesis is possible at weak-scale temperatures if the NLSP decays to LSP sterile sneutrinos.  This asymmetric ``freeze-in" mechanism is more or less identical to the one investigated in \cite{Hall:2010jx} using the superpotential interaction $L_i X H_u$, with $X$ re-interpreted as $N^c_j$.  The lepton asymmetry produced in decays of superpartners to final states involving $\ell_i \tilde{\nu}^c_j$ is proportional to $|\lambda_{ij}|^2 \sin \phi$, where the CP phase $\phi$ arises from soft supersymmetry breaking interactions which might be probed indirectly through electric dipole moments.  As a consequence of \Eq{eq:gammas}, the parameter $|\lambda_{ij}|$ can actually be experimentally accessed via seesaw spectroscopy. 
We leave a more thorough investigation of this remarkable scheme of low-scale leptogenesis for future work \cite{CHP}.

\end{document}